\documentclass[aps,prl,twocolumn,showpacs,superscriptaddress,floatfix]{revtex4}
\usepackage{epsfig}
\newcommand{\be}{\begin{equation}}
\newcommand{\bea}{\begin{eqnarray}}
\newcommand{\eea}{\end{eqnarray}}
\newcommand{\ee}{\end{equation}}

\newcommand{\bra}[1]{\mbox{$\langle #1 |$}}
\newcommand{\ket}[1]{\mbox{$| #1 \rangle$}}

\begin{document}

\title{Stochastic Resonance Phenomena in Quantum Many-Body Systems}

\author {Susana F. Huelga}
\affiliation{Quantum Physics Group, STRI, School of Physics,
Astronomy and Mathematics, \\University of Hertfordshire, Hatfield,
Herts AL10 9AB, UK}
\author{Martin B. Plenio}
\affiliation{QOLS, Blackett Laboratory and Institute for
Mathematical Sciences, \\ 53 Prince's Gate, Exhibition Road,
Imperial College London, London SW7 2PG, UK}
%\maketitle
\date{\today}

\begin{abstract}
We discuss stochastic resonance (SR) effects in driven coupled
quantum systems. We construct dynamical and information theoretic
measures of the system's response that exhibit a non-monotonic
behaviour as a function of the noise strength. We analyze the
relation between lack of monotonicity in the response and the
presence of quantum correlations and identify parameter regimes
where the breakdown of a linear response can be linked to the
presence of entanglement. We also show that a chain of coupled spin
systems can exhibit an array-enhanced response, where the
sensitivity of a single resonator is enhanced as a result of
nearest-neighbour coupling. These results enlarge the domain where
SR effects exist and should be observable in arrays of
superconducting qubits.
\end{abstract}
\pacs{PACS numbers: 03.67.-a, 03.67.Hk} \maketitle
%Introduction.
%
Recently, there has been an increasing interest in the observation
of entanglement and correlation phenomena in arrays of
super-conducting and other solid state qubits realizations
\cite{fazio}. In general, the presence of noise and finite
temperatures is considered detrimental and noise levels are thus
aimed to be minimized. Here, however, we will explore situations
where it is advantageous to maintain a finite, not necessarily
minimal, noise
strength. \\
\noindent The response of an open quantum system to a weak periodic
forcing can exhibit a resonance-like dependence on the noise
strength \cite{qsr}. A canonical example to illustrate this form of
{\em quantum SR (QSR)} is provided by the periodically driven biased
spin-boson model \cite{milena}. Despite quantum coherence was
believed to contribute to the disappearance of SR effects
\cite{rmp}, recent work has shown that QSR should also be displayed
by systems whose dissipative dynamics obeys conventional Bloch
equations \cite{lorenza} and different experimental realizations in
quantum optics have been proposed \cite{bu1}. We will show that
noise-enhanced effects are also present in the steady state response
of quantum-mechanically correlated systems and quantify the presence
of stochastic resonance in terms of both dynamical and information
theoretic
measures.\\
\noindent Our system consists of an array of $N$ driven,
longitudinally coupled spin-1/2 systems (qubits). The system is
subject to a noisy environment modelled by a set of harmonic
oscillators such that each qubit couples transversely to its own
bath.  The global Hamiltonian is given by,
\begin{eqnarray*}
H &=& -\sum_{i=1}^N \frac{\omega_0^i}{2} \, \sigma_z^i + \sum_{k,i}
\omega_k^i (a_{k}^i)^{\dagger} a_k^i + \sum_{i=1}^N \sigma_x^i  X^i
\\ \nonumber &-& \sum_{i=1}^{N-1} J \, \sigma_z^i \otimes
\sigma_z^{i+1}+ \sum_{i=1}^N  \Omega_i \left(\sigma_{+}^i e^{-i
\omega_L^i t} +h.c.\right),
\end{eqnarray*}
where $\hbar=1$, $X^i=\sum_k C_k (a_k^i +a_k^{i \,\dagger})$ denotes
the bath's {\em force operator} and $\sigma_{+}^i=\ket{1}_i
\bra{0}$. The external driving is parameterized by its intensity, as
given by the Rabi frequency $\Omega_i$, and the detuning from the
qubit frequency $\delta_i=\omega_0^i-\omega_L^i$ \cite{jose}. We
will consider situations where the driving is weak and the external
Rabi frequency is smaller than the interqubit coupling, $\Omega < J$
\cite{weak}. Within the rotating wave approximation and for a
Markovian bath, we obtain an effective Hamiltonian for the $N$-qubit
array, $$ H_{\rm eff}= H_{\rm coh}-i\sum_{i=1}^N \Gamma_i
(\bar{n}+1)\, \sigma_{+}^i \sigma_{-}^i-i\sum_{i=1}^N \Gamma_i \,
\bar{n} \,\sigma_{-}^i \sigma_{+}^i, \label{evol}
$$
where,
$
H_{\rm coh}=-\sum_{i=1}^N \frac{\delta_i}{2} \, \sigma_z^i-
\sum_{i=1}^{N-1} J \, \sigma_z^i \otimes \sigma_z^{i+1}+\sum_{i=1}^N
\Omega_i \, \sigma_{x}^i
$
is the coherent part of the Hamiltonian in the interaction picture.
%new
This master equation treatment is valid in the parameter regime
$\Omega_i/\omega \ll 1, \Gamma_i \bar{n}/\omega \ll 1,
\delta_i/\omega \ll 1$ and $J/\omega \ll 1$, where
$\omega=\min\{\omega_0^i, \omega_c\}$ for a suitable frequency cut
off $\omega_c$ of the bath, and all expression in this work will
then be correct to lowest non-trivial order in $\Omega_i, \Gamma_i,
\delta_i
$ and $J$ \cite{cohen}.\\
The noise strength on qubit $i$ at a temperature $T$ is given by the
product $\Gamma_i \, \bar{n}$, where the explicit functional form of
the decay rate $\Gamma_i$ depends on the spectral properties of the
bath and $\bar{n}$ denotes an effective {\em boson} number that
depends on the bath's temperature $T$; both parameters are, in
principle, controllable. For instance, for superconducting qubits,
the parameter $\Gamma$ may be a measure of the fluctuations in the
gate voltages which, if desired, may be amplified on demand
\cite{yuri}. As it will be clear below, our argument is that the
steady state response of the system, as quantified by different
figures of merit, will be optimized at intermediate noise levels and
therefore, trying to reduce the environmental noise to as small as
possible values, does not necessarily provide an optimal universal
strategy to maximize coherent effects.
\begin{figure}
\center{\includegraphics[width=0.65\linewidth]{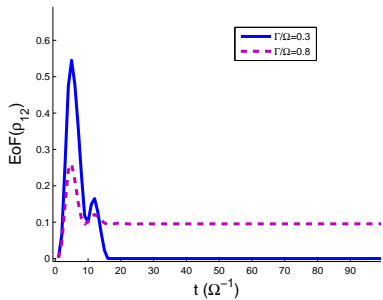}}
%2colourlogo}}
\caption{Entanglement time evolution for two weakly driven qubits
with longitudinal coupling of strength $J/\Omega=1.5$ at $T=0$. If
the noise strength $\Gamma$ is sufficiently large, the system is
inseparable in the steady state (pink dashed line).}
\end{figure}
Let us consider first the case where $N=2$ and $T=0$. By integrating
the time evolution $\dot{\rho}=-i[H_{\rm eff}, \rho]+ \sum_i
\Gamma_i (\bar{n}+1) \, \sigma_{-}^i \rho \sigma_{+}^i +\sum_i
\Gamma_i \bar{n} \, \sigma_{+}^i \rho \sigma_{-}^i$, for fixed
values of the coupling $J$ and the driving
$\Omega=\Omega_1=\Omega_2$, we can analyze whether our system of
weakly driven qubits, initially prepared in their ground state,
develops quantum correlations in time. We employ the entanglement of
formation $E_F$ to quantify bipartite entanglement \cite{eof}. In
Figure 1 we observe that the system will be entangled in the steady
state only for certain finite values of $\Gamma$. Perhaps
surprisingly, it is the largest value of the noise strength the one
that yields steady-state entanglement. %%%%%%%%%%%%%%%%%%%%%%%%%%%%%%%%%%%%%%%%% steady state %%%%%%%%%%%%%%%%%%%%%%%%%%%%%
The steady state of the system can be computed analytically. Calling
$r=\Gamma/\Omega$, $s=J/\Omega$ and $t=r^2+1$,
\begin{equation}
        \rho_{12}^{ss} = \frac{1}{k}\left( \begin{array}{cccc} t^2+4 r^2 s^2 & 2 s r^2+i r t & 2 s r^2+i r t & 2 i r s -r^2\\
                          \ldots & t & r^2 & i r\\
                          \ldots & \ldots & t & i r\\
                          \ldots & \ldots & \ldots & 1\end{array} \right) \; ,\label{ss}
    \end{equation}
where $k=3 +2 r^2 +t^2 + 4r^2 s^2$ and $\ldots$ refer to the
suitable complex conjugate matrix element. The system is entangled,
and have a negative partial transpose, only if $\Gamma>\Gamma_{th}$,
where
\begin{equation}
\Gamma_{th} = \frac{\Omega^2}{2 J}, \label{umbral}
\end{equation}
is the noise threshold. If $\Gamma<\Gamma_{th}$, the state is
separable. This behaviour is illustrated in Figure 2 where the
dashed line corresponds to the bipartite entanglement in the steady
state as quantified by the $E_F$ as a function of the noise strength
$\Gamma$. As a result of the constraint given by Eq.(\ref{umbral}),
any entanglement measure exhibits an initial domain of vanishing
entanglement for weak noise where the state is separable. The
smaller the qubit interaction strength $J$, the larger the value for
the noise required for the driven spins to be entangled. When
$\Gamma$ rises above threshold, the steady state entanglement
increases monotonically up to a maximum at certain optimal noise
strength and decreases steadily for higher values of $\Gamma$
\cite{hans}. This functional form for the bipartite entanglement in
the domain where $\Gamma>\Gamma_{th}$ is reminiscent of stochastic
resonance \cite{qsr} and it had been observed before in the context
of incoherently driven quantum systems \cite{ph}.
\begin{figure}
\center{\includegraphics[width=0.65\linewidth]{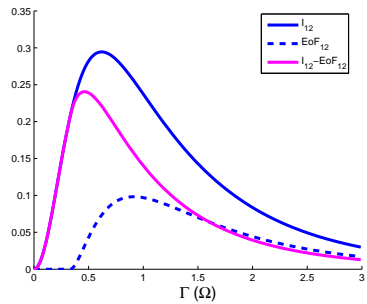}}
%2colourlogo}}
\caption{Stochastic resonance phenomena quantified in terms of
information theoretic measures for a systems of two coupled and
weakly driven spins ($J/\Omega=1.5$) at a zero temperature.
Presented results are for $\delta=0$ but deviations up to
$\delta/\Omega \sim 10^{-2}$ yield very small deviations from the
exact resonance behaviour.}
\end{figure}
%
% True SR
However, to argue for the system to display SR in the conventional
sense, we need to construct suitable measures of the system's
information content and show the characteristic non-monotonic
behaviour as a function of the noise strength that is typical of an
SR response.
%%%%%%%%%%%%%%%%%%%%%%%%% Information Theoretic measure %%%%%%%%%%%%%%%%%%%%%%%%%%%%%%%%%%%%%%%%%%%%%%%%%%%%%%%%%%%%%%%%%%%%%%%%
%
A possible information-theoretic measure is provided by the system's
mutual information $I_{12}= S_{1}+S_{2}-S_{12}$, where $S$ denotes
the von Neumann entropy, $S(\rho)=-tr(\rho \, \log_2 \rho)$
\cite{hanggi}. The blue solid line in Figure 2 is the mutual
information of the steady state $\rho_{12}^{ss}$. Correlations
increase monotonically up to a maximum corresponding to a certain
optimal noise strength above which $I_{12}$ decreases. This
characteristic response is also obtained for the difference between
the mutual information and the entanglement (pink solid line).
Longer chains, as detailed later, also exhibit this type of
non-monotonic response and we therefore argue that SR can be
observed beyond the purely incoherent regime analyzed in
\cite{hanggi}.

%%%%%%%%%%%%%%%%%%%%%%%%%%%%%%%%%%%%%%%%%%%%%%%%%%%%%% Dynamical Measures N=2 %%%%%%%%%%%%%%%%%%%%%%%%%%%%%%%%%%
Alternatively, stochastic resonance can be characterized using a
dynamical measure of the system's response to the external driving.
In the case of single quantum systems, it has been proposed to use
as an appropriate figure of merit the expectation value of a
suitable Pauli operator \cite{nancy}.
% Mention generalization of quantities used by Nancy and Buchleitner
This suggests considering local observables of the form $\cal{
S}$$_{N}=\langle(1/N)\sum_{i=1}^N \sigma_{\xi}^{i}\rangle$
$(\xi=x,z)$ to characterize the system's dynamical response in the
multipartite scenario. Note that $\cal{S}$$_N$ is easily accessible
experimentally and may be even measured without local control.\\ We
can evaluate analytically the form of the signal
$\cal{S}$$_2=\langle(\sigma_{x}^1 + \sigma_{x}^2)/2\rangle=4
sr^2/k$, which is non-monotonic as a function of both the coupling
$J$ between qubits and the coupling to the reservoir, reaching a
maximum at certain intermediate values. The signal $\cal{S}$$_2$ is
maximum for $\Gamma= \sqrt{2} \Omega$, independently of $J$, with
$\cal S$$_{2,\rm max}=s/(2+s^2)$. The amplitude of the signal is
maximal for $s=J/\Omega=\sqrt{2}$ and as the ratio $s$ increases,
the response becomes weaker and loses its resonance-like shape,
resembling the shape of a saturation curve .
%%%%%%%%%%%%%%%%%%%%%% Relation monotonicy-entanglement %%%%%%%%%%%%%%%%%%%%%%%%%%%%%%%%%%%%%%%%%5
In general, for arbitrary values of $\Omega$ and $J$, the
transition from linear to non-linear response has no direct
relation with the presence of quantum correlations in the system.
The system will be separable in the linear region if $\Gamma_{th}
> \sqrt{2} \Omega $ while it can be entangled and responding
linearly if $\Gamma_{th}<\Gamma <\sqrt{2} \Omega$. However, tuning
the nearest neighbour coupling $J$ to the value $\Omega/ \sqrt{8}$
yields a maximal response at exactly the transition point from PPT
$\mapsto$ NPT states, i.e. from separable to entangled subsystems.
As a result, if $\Omega<\sqrt{8} J$, we can ensure that the
breakdown of the monotonic response when varying $\Gamma$ implies
the presence of entanglement in the system. This happens in the
regime where $\Omega<J$ for both the dynamical response
$\cal{S}$$_2$ and the SR information theoretic measures: The maximum
values of the response are reached for a noise strength above
threshold and the lack of monotonicity allows to conclude that there
is entanglement in the system. Note, however, that the opposite
conclusion, i.e. monotonicity $\Longrightarrow$ separability, is not
correct in
general, even for weak driving.\\
\begin{figure}
\center{\includegraphics[width=0.65\linewidth]{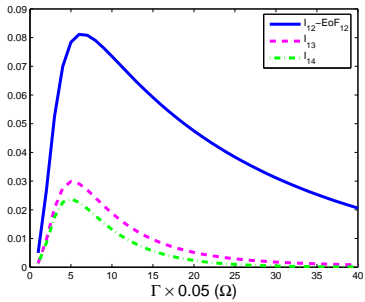}}
%2colourlogo}}
\caption{System's response for a chain of $N=4$ qubits at $T=0$ with
$J_i/\Omega_i=1.5$, $(i=1 \dots 4)$ as quantified by an information
theoretic measure, the quantum mutual information. Under the
evolution given by $H_{\rm eff}$, entanglement is restricted to
nearest neighbours only and $E_{F_{13}}=E_{F_{14}}=0$.}
\end{figure}
%
%%%%%%%%%%%%%%%%%%%%%%%%%%%%% Chain of four-five qubits %%%%%%%%%%%%%%%%%%%%%%%%%%%%%%%%%%%%%%%%%%%%%%%%%%%
%
SR phenomena, as quantified by information theoretic and dynamical
measures, should also be observable in chains of longitudinally
coupled weakly driven spin systems. With the type of interaction we
have considered, quantum correlations are confined to nearest
neighbours and retain the same qualitative behaviour discussed in
the $N=2$ case, with quantum correlations only arising above a noise
threshold that decreases as function of $N$. In Figure 3 we have
characterized the system's response for a chain with $N=4$ in terms
of the quantum mutual information between qubits 1 and $j$,
$(j=2,3,4)$. We observe a resonance-like response as a function of
the noise strength at $T=0$ for both nearest and distant neighbours.
%
%%%%%%%%%%%%%%%%%%%%%%%%%%%%%%%%%%%%%%%%%%%%%%%% Dependence with T %%%%%%%%%%%%%%%%%%%%%%%%%%%%
So far we have considered the environment to be at zero temperature.
All phenomena described so far are robust in the presence of a
finite $T$, with the net result that maximum values in both
information theoretic and dynamical measures of SR are reduced with
increasing $T$ (See Figure 5 as an illustration for typical
superconducting qubits operating temperatures). The mutual
information $I_{ij}$ decreases monotonically with $\bar{n}$, as
illustrated by the blue solid line in Figure 4 for qubits 1 and 2 in
a chain with $N=4$, and so does the steady state entanglement, which
becomes zero for $T$ sufficiently high. We should stress that
$I_{12}$ quantifies 'total' correlations, not distinguishing between
quantum and classical contributions. Given that $I_{ij}$ measures
total correlations, and the $E_F$ is a measure of entanglement(but
not the only one), their difference $I_{ij}-E_{F{ij}}$ may be viewed
as a reasonable attempt to measure the classical correlation content
of a quantum state. This quantity displays the characteristic SR
shape as $T$ increases, with the maximum response being reached at
the point where $E_{F_{12}}=0$ (pink solid line in Figure 4)
\cite{finiteT}.

\begin{figure}[ht]
\begin{minipage}{0.23\textwidth}
\includegraphics[height=2.9cm]{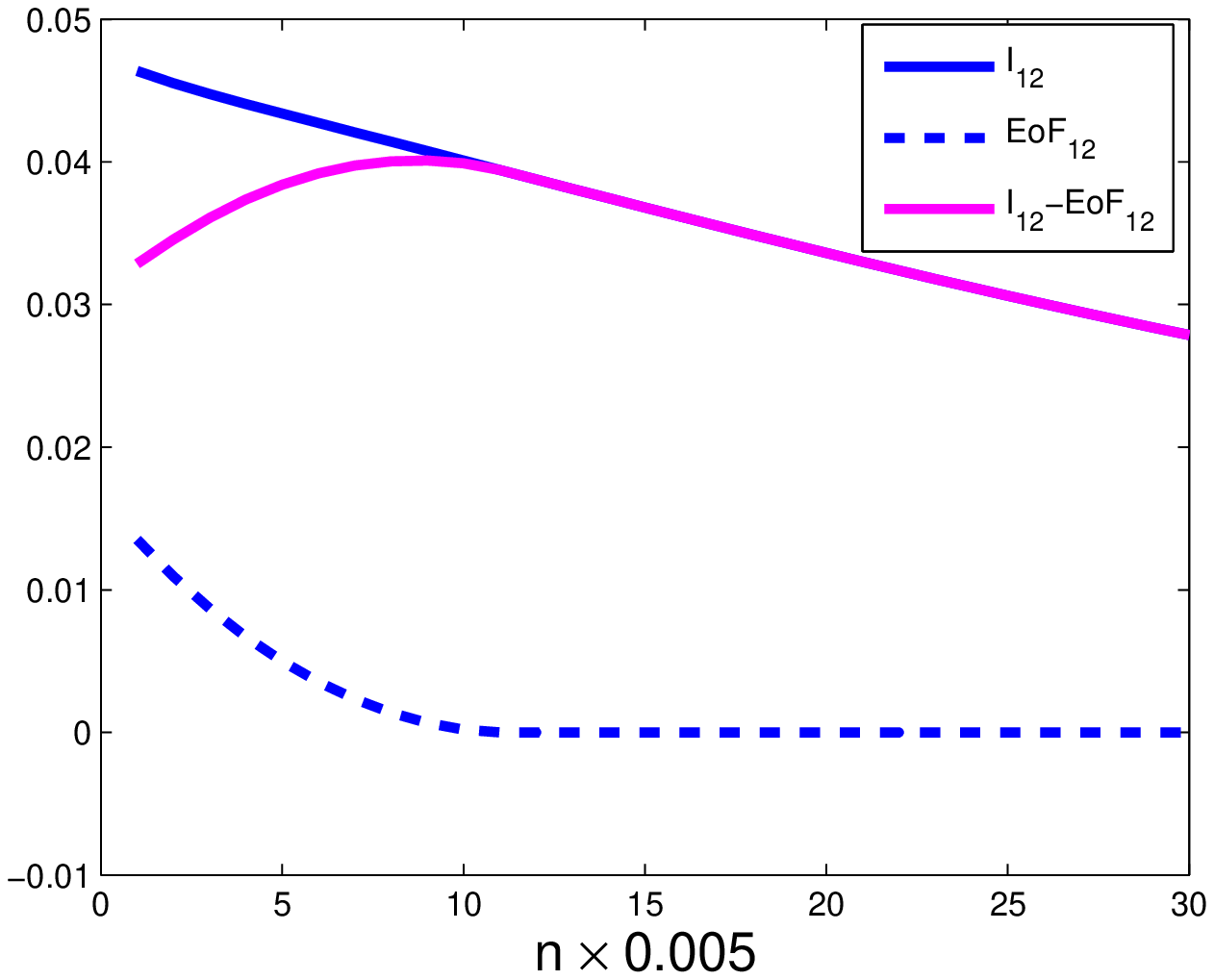}
\caption{Information theoretic measures as a function of the
bath's mean boson number (temperature) for a chain of N=4 spin
with $J_i/\Omega_i=1.5$ and $\Gamma_i/\Omega_i=1$, $i=1 \dots 4$.}
\end{minipage}\hspace{2mm}%
\begin{minipage}{0.23\textwidth}
\includegraphics[height=2.9cm]{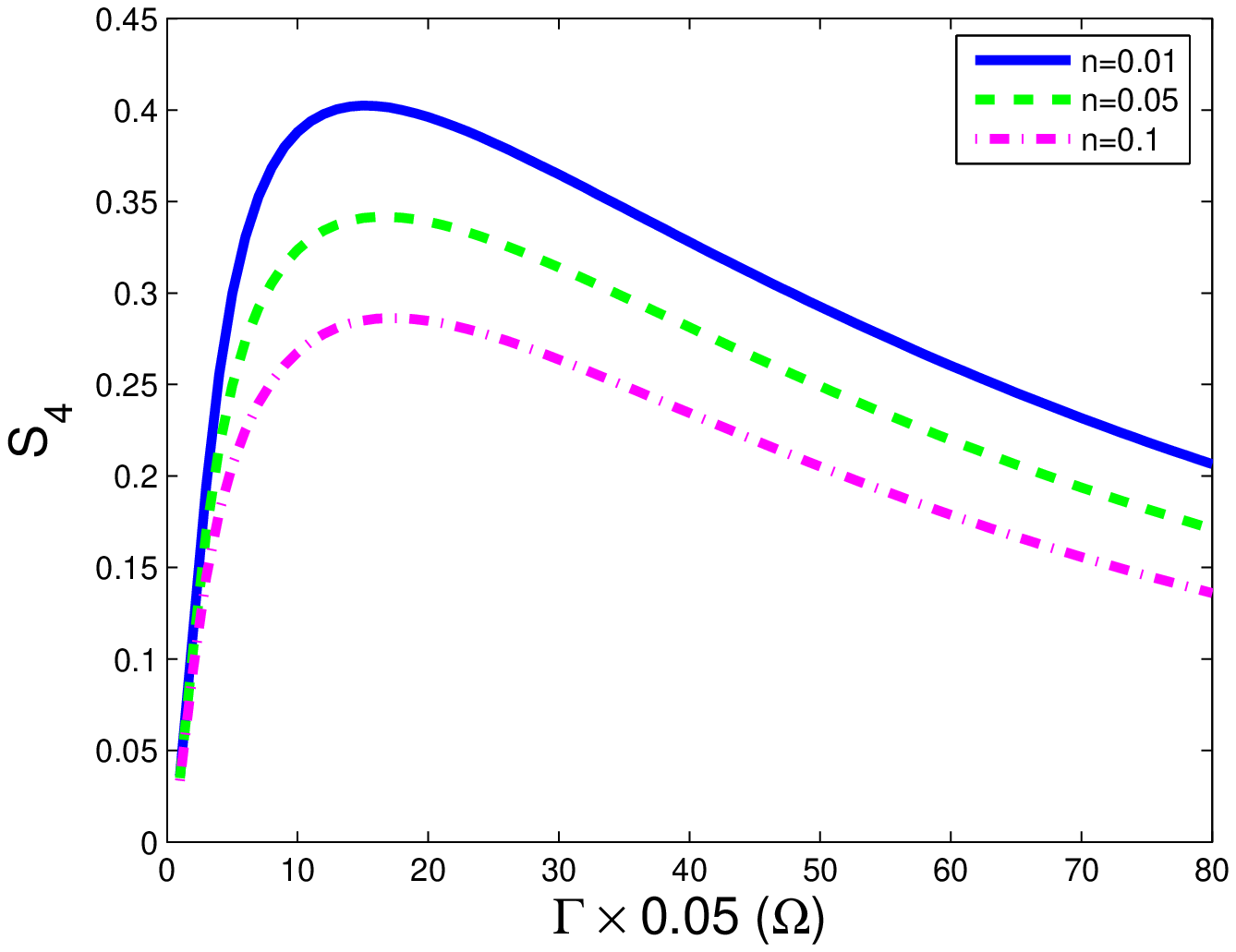}
\caption{System's response for a chain of $N=4$ qubits with
$J/\Omega=1.5$ as quantified by $\cal{S} $$_4$$=\langle
\sum_{i=1}^4 \sigma_x^i/4 \rangle$ as a function of the noise
strength $\Gamma$ and for different values of the external
temperature.}
\end{minipage}

\end{figure}

%%%%%%%%%%%%%%%%%%%%%%%%%%%%%%%%%%% Chain Dynamical response  %%%%%%%%%%%%%%%%%%%%%%%%%%%%%%%%%%%%%%%%%%%%%%%%%%% Conclusions %%%%%%%%%%%%%%%%%%%%%%
The non-monotonicity of the response is also apparent in the
dynamical measure $\cal S$$_4$, plotted in Figure 5 for increasing
values of the mean thermal boson number $\bar{n}$. At a given $T$,
the value of $\Gamma$ that maximizes the response is now a function
of both $\Omega$ and $J$ and a numerical analysis shows that we can
link entanglement and lack of monotonicity whenever outside the
regime where $\Omega \gg J$.
It is interesting to compare this global response with the
individual signal that can be obtained from each qubit alone.
Classical stochastic resonators are known to display an
array-enhanced SR, where the collective dynamics yields the
amplification of some suitably defined signal-to-noise ratio of a
single oscillator \cite{jung, adi}. On exact resonance, the steady
state expectation value $\langle \sigma_x \rangle$ for a driven
isolated qubit transversely coupled to a bosonic environment at zero
temperature is strictly zero, given that $\langle \sigma_x \rangle =
\rho_{01}+\rho_{10} \sim \Omega \delta$. For a finite detuning, the
single qubit response is a monotonically decreasing function of the
noise strength. Specifically, $\mid \langle \sigma_x
\rangle_{ss}\mid = \Omega \, \delta/ (\delta^2 +(\Gamma/2)^2
+(\Omega/2)^2)$. Introducing interqubit coupling leads to a build up
of a real part in the single qubit coherences which is proportional
to the coupling strength $J$ to first order. For $N=2$, $\langle
\sigma_x \rangle = \frac{4 J \,\Gamma^2}{ k \, \Omega^3}$ and the
single qubit response displays the typical SR profile, while
increasing the size of the array yields a sharper signal with a
maximum value that increases as a function of $N$ and that is
obtained for increasingly smaller values of the noise parameter, as
noted in Figure 6.
\begin{figure}
\center{\includegraphics[width=0.65\linewidth]{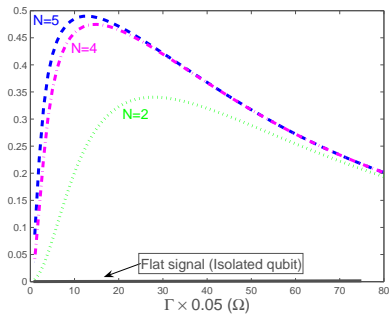}}
%2colourlogo}}
\caption{Single qubit steady state response for weakly driven
arrays of 2, 4 and 5 qubits with $J/\Omega=1.5$ and
$\bar{n}=0.01$. The zero signal corresponds to a resonantly driven
uncoupled oscillator. Interqubit coupling yields an enhanced
response as a function of the number $N$ of oscillators in the
array.}
\end{figure}
The dashed-dotted line corresponds to the signal for the first qubit
in an array of $N=4$, while the dashed line corresponds to the
response of the first qubit alone in an array of 5 spins. If we
compare the numerical value of the single qubit expectation values
with the global response specified in Figure 5, we note that the
single qubit response is enhanced as a result of the coupling, which
shows the persistence of array enhancement effects
in chains of quantum spin systems.\\
%%%%%%%%%%%%%%%%% Array enhancement %%%%%%%%%%%%%%%%%%%%%%%%%%%%%%%%%%%%%%%%%%%%%%%
It seems clear from the present analysis that stochastic resonance
phenomena predicted for coupled classical resonators, including
those obeying an Ising model \cite{ising}, may also be observed
within coupled qubits and coexist with the presence quantum
correlations. These results are amenable to experimental
verification on a variety of proposed qubit realizations \cite{new}.
Solid state architectures, and in particular superconducting qubits
\cite{yuri}, combine noisy environments with a degree of
qubit-environment tunability that made them particularly suited to
demonstrate SR in the terms discussed in this Letter.
\\

%----------------------------------------------------------
\acknowledgements \noindent We are grateful to A. Bulsara, M.
Dykman, T. Clark, J. Ralph and D. Tsomokos for useful discussions
and to the EPSRC QIP-IRC, the EU Integrated Project QAP, and the
Royal Society for support.


\begin{thebibliography}{00}
\bibitem{jose}
This dynamics may describe the evolution of an array of
superconducting qubits operated at the degeneracy point with $XX$
nearest neighbour coupling and under the action of gaussian noise
due to the surrounding circuitry.
\bibitem{weak}
Outside this regime, the system is separable for most noise
strenghts; moreover the resonance-like shape of the proposed
figures of merit to characterize SR becomes much less well
defined.
\bibitem{finiteT}
Parameter regimes can be identified where the mutual information,
both for nearest and distant neighbours, displays SR as a function
of $\bar{n}$. In those situations we encounter the nearest neighbour
entanglement to be strictly zero.
\bibitem{new} We have examples that demonstrate that our results persist beyond
the Born-Markov framework. A detailed analysis of QSR in
multipartite systems subject to a non-Markov environment is
currently underway and will be presented in a forthcoming paper.
\end{thebibliography}
\end{document}